\documentclass[fleqn,usenatbib]{mnras}

\usepackage{newtxtext,newtxmath}

\usepackage[T1]{fontenc}
\usepackage{ae,aecompl}

\usepackage{graphicx}	
\usepackage{amsmath}	
\usepackage{amssymb}	

\title[Testing asteroseismic radii of dwarfs]{Asteroseismic radii of dwarfs: New accuracy constraints from \textit{Gaia} DR2 parallaxes}

\author[C. L. Sahlholdt \& V. Silva Aguirre]{
Christian L. Sahlholdt$^{1}$\thanks{E-mail: sahlholdt@astro.lu.se},
Victor Silva Aguirre$^{2}$
\\
$^{1}$Lund Observatory, Department of Astronomy and Theoretical Physics, Box 43, SE-221 00 Lund, Sweden\\
$^{2}$Stellar Astrophysics Centre, Department of Physics and Astronomy, Aarhus University, Ny Munkegade 120, DK-8000 Aarhus C, Denmark\\
}

\date{Accepted XXX. Received YYY; in original form ZZZ}

\pubyear{2018}

\begin{document}
\label{firstpage}
\pagerange{\pageref{firstpage}--\pageref{lastpage}}
\maketitle

\begin{abstract}
Precise stellar masses and radii can be determined using asteroseismology, but their accuracy must be tested against independent estimates.
Using radii derived from \textit{Gaia} DR2 parallaxes, we test the accuracy of asteroseismic radii for a sample of 93 dwarfs based on both individual frequency fitting and the seismic scaling relations.
Radii from frequency fitting are about 1 per cent smaller than \textit{Gaia} radii on average; however, this difference may be explained by a negative bias of $30~\mathrm{\upmu as}$ in the \textit{Gaia} parallaxes.
This indicates that the radii derived from frequency fitting are accurate to within 1 per cent.
The scaling relations are found to overestimate radii by more than 5 per cent, compared to the \textit{Gaia} radii, at the highest temperatures.
We demonstrate that this offset is reduced to 3 per cent after applying corrections based on model frequencies to the scaling relation for $\Delta\nu$, but only when the model frequencies are corrected for the surface effect.
With corrections to $\Delta\nu$, the scaling relation gives radii accurate to about 2--3 per cent for dwarfs in the temperature range $5400$--$6700$~K.
The remaining offset at the highest temperatures may indicate the need for a correction to the scaling relation for $\nu_{\mathrm{max}}$.
\end{abstract}

\begin{keywords}
asteroseismology -- stars: fundamental parameters
\end{keywords}



\section{Introduction}
During the last few decades, asteroseismology of solar-like oscillators has emerged as a powerful tool for determining precise stellar parameters, not least thanks to the high-precision photometry from space-based telescopes (see \citet{2013ARA&A..51..353C} for a review).
In the best-case scenario, the frequencies of the individual stellar oscillations can be identified in the power spectrum and fitted to stellar evolutionary models in order to determine precise fundamental parameters like the radius, mass, age, etc. \citep[e.g.][]{2015MNRAS.452.2127S, 2017ApJ...835..173S}.
However, for the large majority of targets observed by the space missions, the signal-to-noise ratio is not high enough to reliably identify the individual frequencies.
Still, two average parameters of the power spectrum can be determined: the mean frequency separation, $\Delta\nu$, and the frequency of maximum power, $\nu_\mathrm{max}$.
These parameters scale with the stellar mean density and surface gravity, and allow for estimates of the stellar mass and radius through a set of scaling relations.
The scaling relations have been applied to estimate stellar parameters for more than 500 main sequence and subgiant stars \citep{2014ApJS..210....1C} and thousands of red giants \citep[e.g.,][]{2018arXiv180409983P,2018MNRAS.475.5487S} observed by the \textit{Kepler} spacecraft .

In recent years, much effort has been devoted to testing the accuracy of the stellar parameters derived from asteroseismic analyses.
For the scaling relations, such investigations have shown that the seismic radii of dwarfs are accurate to within about 5 per cent, based e.g. on comparisons with radii from interferometry \citep{2012ApJ...760...32H, 2013MNRAS.433.1262W}.
Following the first data release from the \textit{Gaia} mission (\textit{Gaia} DR1; \citealt{2016A&A...595A...1G, 2016A&A...595A...2G}), a number of studies were carried out comparing distances (or parallaxes) derived from seismic radii from scaling relations to those from the Tycho-\textit{Gaia} astrometric solution \citep{2016A&A...595A...4L}.
These studies generally found good agreement between seismic and \textit{Gaia} parallaxes for dwarf stars, confirming once again that the scaling relations provide radii accurate to within 5 per cent \citep{2016A&A...595L...3D, 2017ApJ...844..102H}.
However, seismic parallaxes derived from the radii of dwarfs based on modelling of individual frequencies have been found to be systematically overestimated by about 3 per cent compared to the \textit{Gaia} DR1 parallaxes \citep{2017ApJ...835..173S, 2018MNRAS.476.1931S}.
Due to the possibility of systematic errors in the \textit{Gaia} DR1 parallaxes, it was not clear whether this was due to a 3 per cent systematic in the seismic radii or the \textit{Gaia} parallaxes (or a combination of the two).

With the recent second data release from the \textit{Gaia} mission (\textit{Gaia} DR2; \citealt{2018arXiv180409365G, 2018arXiv180409366L}) the precision of the parallaxes of \textit{Kepler} stars has improved significantly and the level of systematic errors has decreased.
In this paper we use parallaxes from \textit{Gaia} DR2, to put improved constraints on the accuracy of seismic radii of dwarf stars\footnote{Most of our analysis has been carried out in a Jupyter notebook which is publicly available on GitHub: \url{https://github.com/csahlholdt/testing_ast_radii}.}.

\begin{figure}
  \center
  \includegraphics[width=\columnwidth]{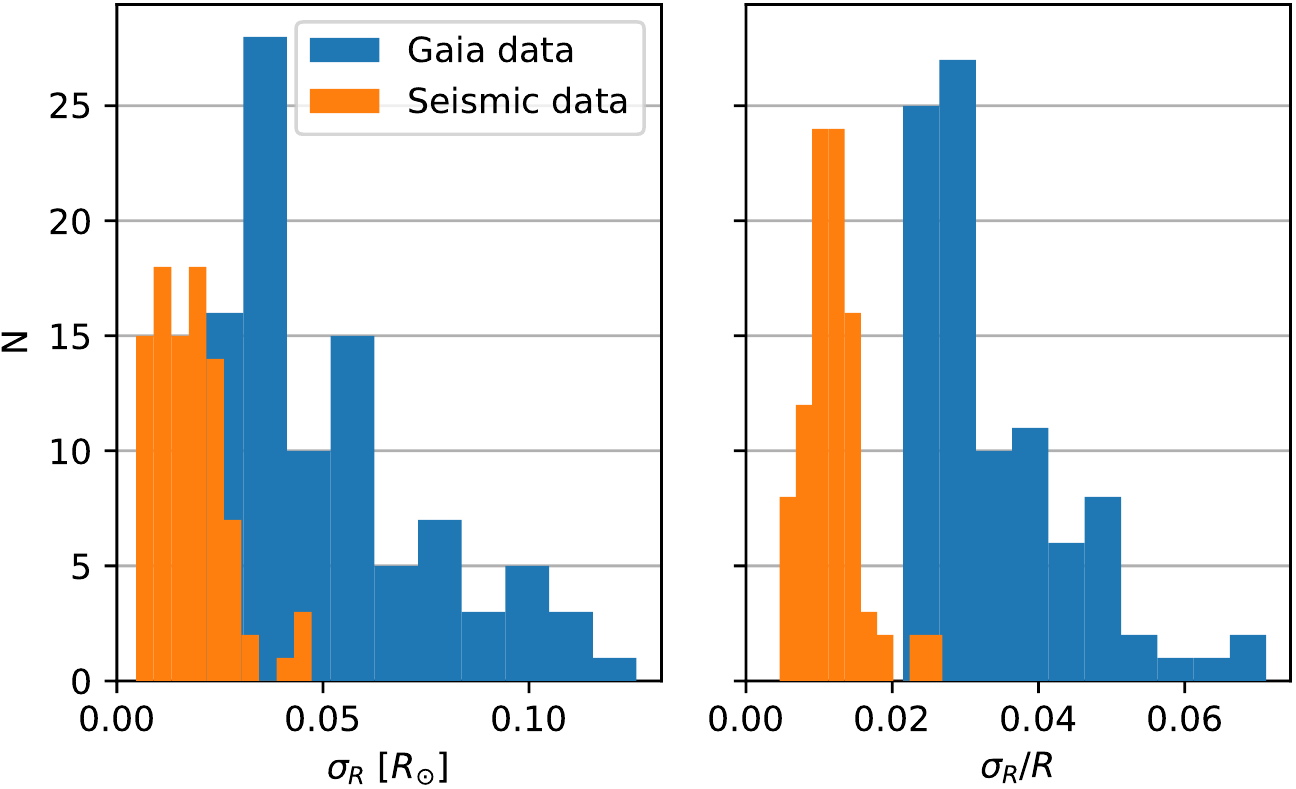}
  \caption{Distributions of absolute (left) and relative (right) uncertainties on the radii determined from \textit{Gaia} parallaxes and from asteroseismology using individual frequency modelling.}
  \label{fig:rad_hist}
\end{figure}

\section{Sample and Methods}
We consider a sample of dwarf stars with individual frequencies, as well as $\Delta\nu$ and $\nu_\mathrm{max}$, determined from \textit{Kepler} data.
The sample consists of the 33 stars of `simple' type from the Kages sample \citep{2016MNRAS.456.2183D} and the 66 stars of the LEGACY sample \citep{2017ApJ...835..172L}.
These two studies had 4 stars in common making the total sample size 95 stars.
Two of these were not included in \textit{Gaia} DR2 which leaves us with 93 dwarfs with individual frequencies and \textit{Gaia} parallaxes.
The typical parallax precision for these stars has improved by a factor of 10 from $300~\mathrm{\upmu as}$ in \textit{Gaia} DR1 to $30~\mathrm{\upmu as}$ in \textit{Gaia} DR2.

We have collected spectroscopic metallicities, [Fe/H], for the sample from \citet{2015ApJ...808..187B}.
Some of the LEGACY stars were not included in that study, for these we use the same [Fe/H] as in the original study of the LEGACY stars \citep{2017ApJ...835..173S}.
For effective temperatures, $T_{\mathrm{eff}}$, and angular diameters, $\theta$, we adopt the values based on the infrared flux method (IRFM; calibration by \citealt{2010A&A...512A..54C}) derived in our previous study of this sample \citep{2018MNRAS.476.1931S}.
The IRFM angular diameters are derived from the fundamental relations between bolometric flux, effective temperature, and angular diameter (see \citealt{2006MNRAS.373...13C}); hence, they correspond to limb-darkening corrected angular diameters obtained from interferometry.
They are, however, subject to uncertainties related to bolometric corrections, reddening, and the zero-point of the temperature scale.
Throughout our analysis we assume that there is no significant bias in the angular diameters.

\subsection{Seismic radii}
We test two sets of seismic radii: one calculated by fitting  stellar models to the observed frequencies, and the other calculated directly from the seismic scaling relations.
For the first set we use stellar models calculated using the Garching Stellar Evolution Code (\verb|GARSTEC|; \citealt{2008Ap&SS.316...99W}) with theoretical oscillation frequencies from the Aarhus Adiabatic Oscillation Package (\verb|adipls|; \citealt{2008Ap&SS.316..113C}).
When fitting to the frequencies, we use the ratios of frequency differences described by \citet{2003A&A...411..215R}.
The advantage is that these ratios are insensitive to the stellar surface layers, so the systematic overestimation of the highest model frequencies, known as the surface effect, becomes irrelevant.
For simplicity, we refer to this as fitting to individual frequencies.
We fit the stellar models to the individual frequencies, $T_{\mathrm{eff}}$, and [Fe/H] using the latest version of the BAyesian STellar Algorithm (\verb|BASTA|; \citealt{2015MNRAS.452.2127S}).
The stellar models are the same as those described in \citet[section 3.1]{2018MNRAS.476.1931S}.

For the second set of seismic radii we use the scaling relations for $\Delta\nu$ and $\nu_\mathrm{max}$ which are given by \citep{1986ApJ...306L..37U, 1991ApJ...368..599B}
\begin{align}
  \frac{\Delta\nu}{\Delta\nu_{\sun}} &\simeq
  \left(\frac{M}{M_{\sun}}\right)^{1/2}
  \left(\frac{R}{R_{\sun}}\right)^{-3/2} \label{eq:dnu_scal} \; ,
  \\
  \frac{\nu_{\mathrm{max}}}{\nu_{\mathrm{max},\sun}} &\simeq
                          \left(\frac{M}{M_{\sun}}\right)
                          \left(\frac{R}{R_{\sun}}\right)^{-2}
                          \left(\frac{T_{\mathrm{eff}}}
                          {T_{\mathrm{eff},\sun}}\right)^{-1/2}
                          \label{eq:numax_scal} \; .
\end{align}
Combining these relations, the radius is given by
\begin{align}
  \frac{R}{R_{\sun}} &\simeq
  \left(\frac{\nu_{\mathrm{max}}}{\nu_{\mathrm{max},\sun}}\right)
  \left(\frac{\Delta\nu}{\Delta\nu_{\sun}}\right)^{-2}
  \left(\frac{T_{\mathrm{eff}}}{T_{\mathrm{eff},\sun}}\right)^{1/2}
                              \; . \label{eq:R_scal}
\end{align}
For the solar values we adopt ${\nu_{\mathrm{max,\sun}} = 3090~\mathrm{\upmu Hz}}$, ${\Delta\nu_{\sun} = 135.1~\mathrm{\upmu Hz}}$ \citep{2011ApJ...743..143H}, and ${T_{\mathrm{eff,\sun}} = 5777}$~K.

\subsection{\textit{Gaia} radii}
To calculate a set of radii based on the \textit{Gaia} parallaxes, $\varpi$, we use them in combination with the IRFM angular diameters.
The stellar radius and angular diameter can be used to calculate the distance and thereby the parallax:
\begin{equation} \label{eq:dist_par}
  d = C\frac{2R}{\theta} \;\;\; \Rightarrow \;\;\;
  \varpi = \frac{1}{d} = \frac{\theta}{C\times 2R} \; ,
\end{equation}
where $C$ is  the factor which converts the distance into parsec.
Then the radius is given by
\begin{equation}
  R = \frac{\theta}{C\times 2\varpi} \; .
\end{equation}
In \autoref{eq:dist_par} we assume that the inverse of the parallax is a good estimate of the distance which is not necessarily the case when the parallax is very uncertain (see e.g. \citealt{2015PASP..127..994B}).
However, for all but two of the stars in our sample, the parallax uncertainties are below 2 per cent (the other two have uncertainties of 3 and 5 per cent), so we do not expect this assumption to bias our radii to any significant degree.

In \autoref{fig:rad_hist} we show the uncertainties on the \textit{Gaia} radii and the seismic radii derived from individual frequency fitting.
The seismic radii all have uncertainties below $0.05~R_{\odot}$ or about 2 per cent.
Overall, the \textit{Gaia} radii are more uncertain, and the majority of them are precise to within 5 per cent.
The precision of the \textit{Gaia} radii is limited by the angular diameters which have uncertainties of about 3--5 per cent.
This includes uncertainties in the photometry, reddening, temperature, metallicity, and surface gravity, and these terms are now the limiting factors, rather than the parallax, in the radius precision for these stars.

\begin{figure*}
  \center
  \includegraphics[width=0.9\textwidth]{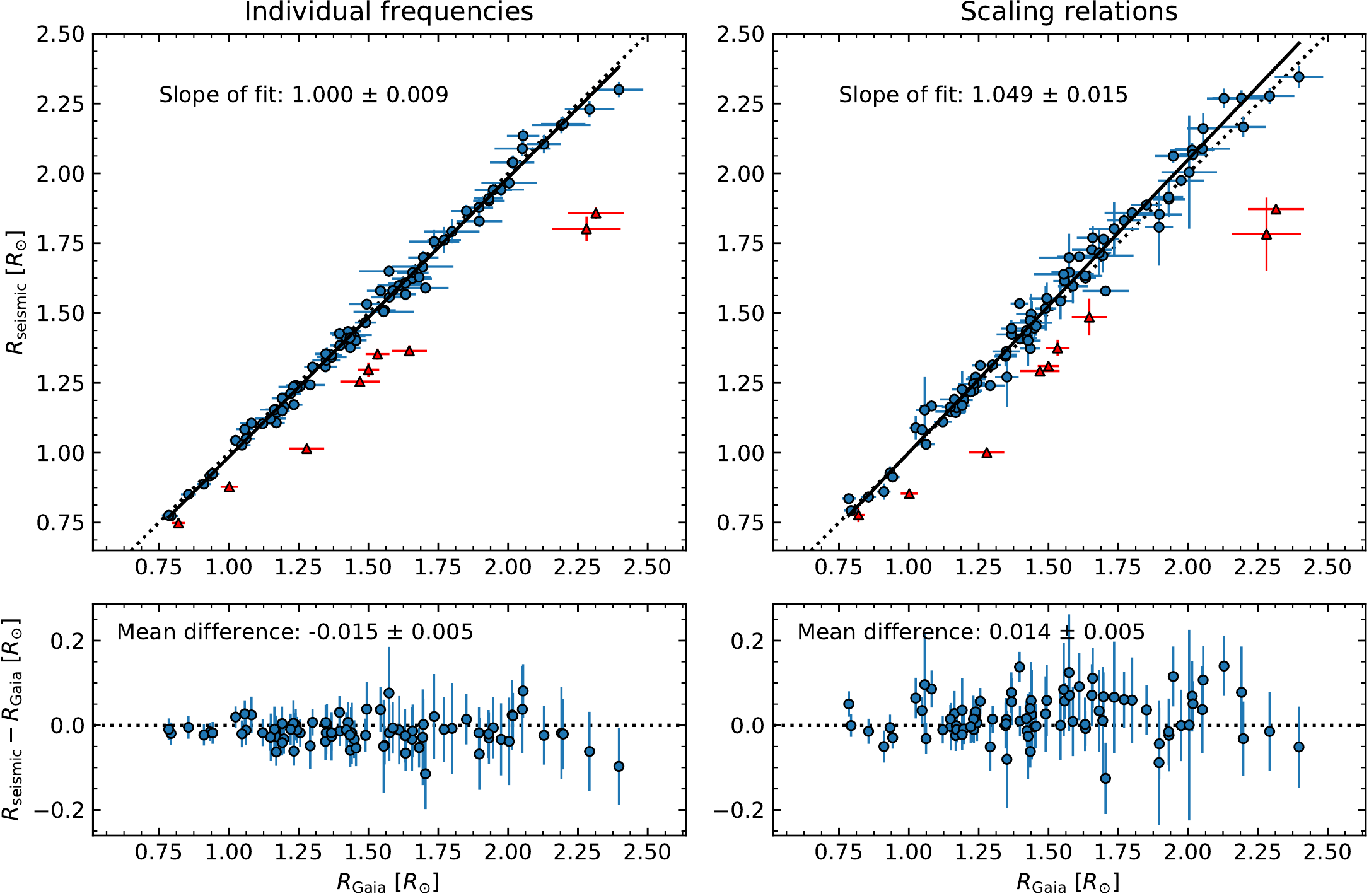}
  \caption{Comparison between \textit{Gaia} and seismic radii for the analyses based on modelling individual frequencies (left) and scaling relations (right).
  In the upper panels, the points marked in red are those for which the radius difference is larger than three standard deviations.
  These outliers have been identified based on the individual frequency analysis and are the same in both panels.
  The dotted line is the 1:1 relation and the solid line is a linear fit using orthogonal distance regression (as implemented in the python package \texttt{scipy.odr}) including uncertainties for weighting.
  In the lower panels, the radius differences are shown for all of the non-outliers.
  The dotted line marks zero difference, and the mean differences given in the plot are calculated with each point weighted according to its uncertainty.}
  \label{fig:rad_comp}
\end{figure*}

\begin{figure}
  \center
  \includegraphics[width=\columnwidth]{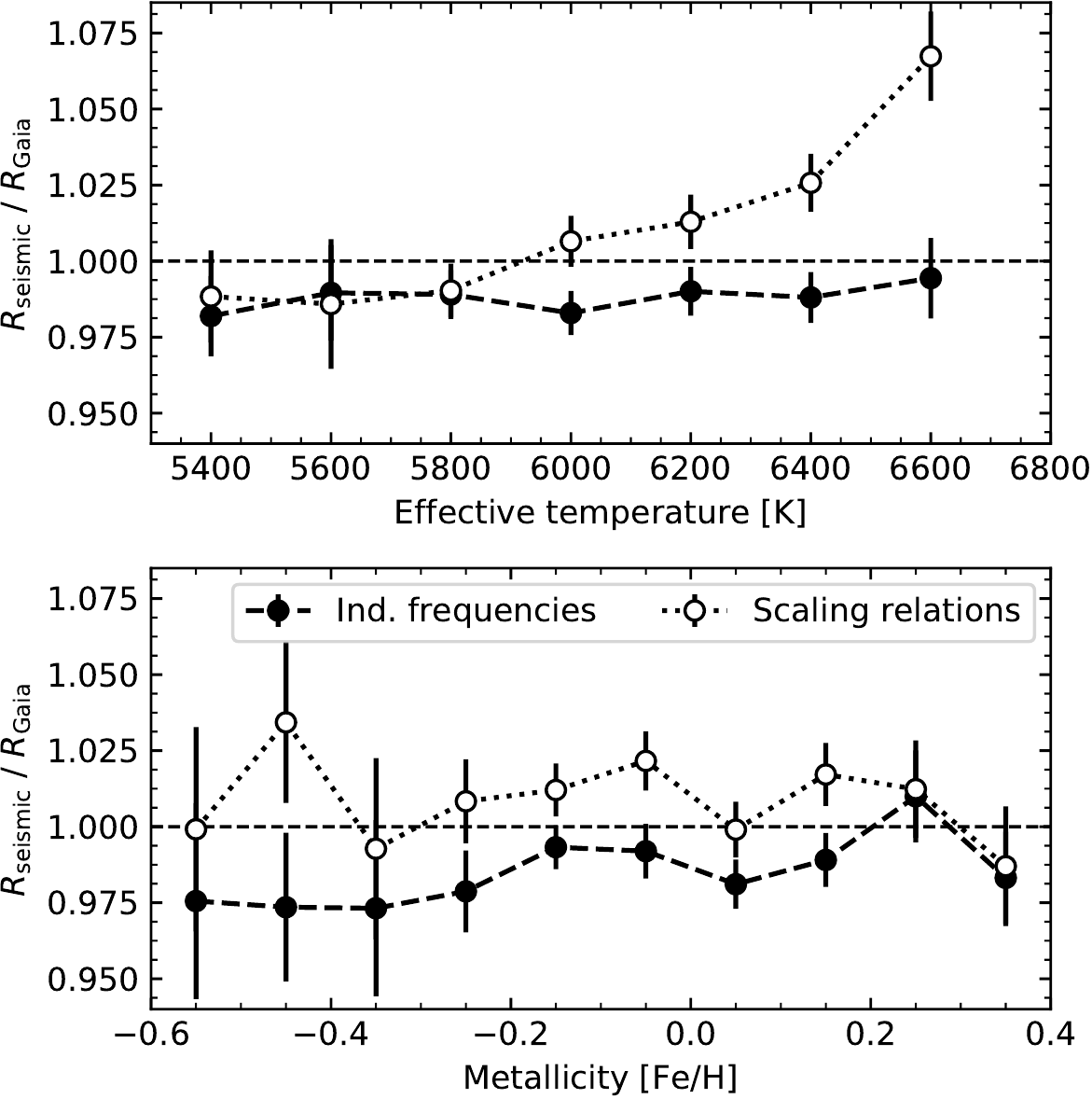}
  \caption{Mean ratios of seismic to \textit{Gaia} radii binned by the effective temperature (top) and the metallicity (bottom).
  The filled circles show the comparison with seismic radii based on modelling individual frequencies and the open circles are based on the scaling relations.}
  \label{fig:rad_ratio}
\end{figure}

\begin{figure}
  \center
  \includegraphics[width=\columnwidth]{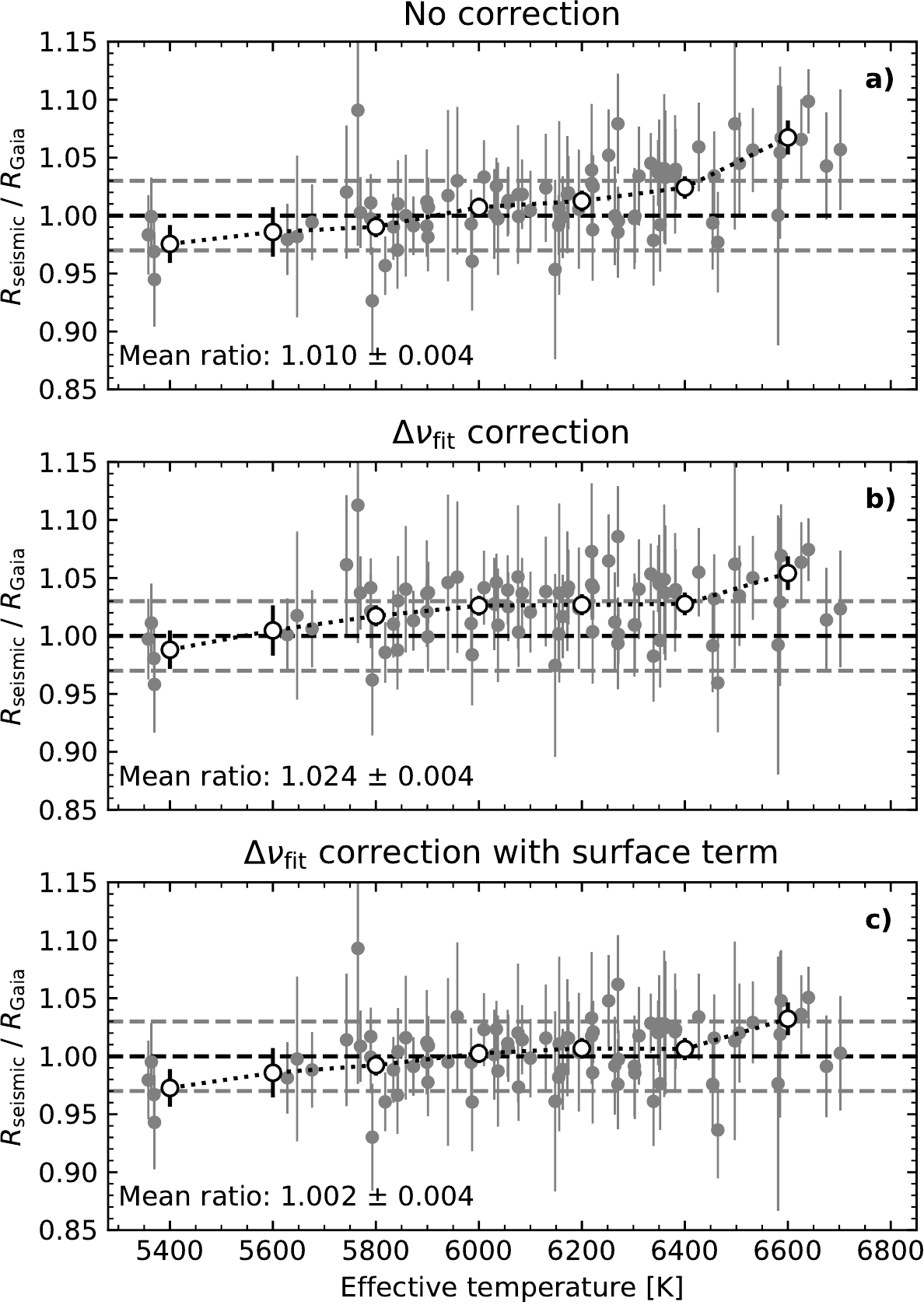}
  \caption{Ratios of seismic to \textit{Gaia} radii as a function of temperature, with seismic radii based on the scaling relations.
  The black dashed line marks unity, and the grey dashed lines mark a difference of 3 per cent to guide the eye.
  The open circles are mean values binned by temperature.
  \textbf{a)} No corrections to the scaling relation (i.e. \autoref{eq:R_scal}).
  \textbf{b)} $\Delta\nu$ corrected using $\Delta\nu_{\mathrm{fit}}$ from the models of our best fits to individual frequencies.
  \textbf{c)} Like \textbf{b)} but with $\Delta\nu_{\mathrm{fit}}$ corrected for the surface effect.
  See the text for more details on the $\Delta\nu_{\mathrm{fit}}$ correction.}
  \label{fig:rad_ratio_fit}
\end{figure}

\section{Results and discussion}
In \autoref{fig:rad_comp}, the two sets of seismic radii are compared with the \textit{Gaia} radii for the full sample of 93 stars.
A number of outliers are clearly visible; we have marked in red the stars for which the difference between the \textit{Gaia} radius and the seismic radius based on individual frequency fitting is greater than $3\sigma$.
They are all outliers in the sense that the \textit{Gaia} radius is too large which may be explained by contaminated photometry leading to overestimated angular diameters.
In all results discussed in the following, these nine outliers have been removed leaving 84 stars.

By fitting a straight line to the data using orthogonal distance regression, we find that the slope is consistent with one for radii based on individual frequencies.
However, there is a small but significant average offset of ${\langle R_{\mathrm{seismic}}-R_{\mathrm{Gaia}}\rangle = -0.015\pm 0.005~R_{\odot}}$ suggesting that the seismic radii are underestimated.
If we make the comparison in terms of parallaxes instead, by calculating seismic parallaxes using \autoref{eq:dist_par}, we find a mean difference of ${\langle\varpi_{\mathrm{Gaia}}-\varpi_{\mathrm{seismic}}\rangle = -35\pm 16~\mathrm{\upmu as}}$ with seismic parallaxes based on the individual frequency analysis.
This offset is similar in direction and magnitude to those found based on seismic parallaxes of red giants in the \textit{Kepler} field \citep{2018arXiv180502650Z} and parallaxes of eclipsing binaries \citep{2018arXiv180503526S}.
Our offset is also in agreement with the possible negative bias in the \textit{Gaia} DR2 parallaxes of about $30~\mathrm{\upmu as}$ based on the parallaxes of quasars \citep{2018arXiv180409366L}.
This means that the offset between seismic and \textit{Gaia} radii in the left-hand panels of \autoref{fig:rad_comp} may be entirely explained by a negative bias in the \textit{Gaia} parallaxes leading to overestimated \textit{Gaia} radii.
For the comparison based on scaling relations, we find a slope larger than one and an average offset of ${\langle R_{\mathrm{seismic}}-R_{\mathrm{Gaia}}\rangle = 0.014\pm 0.005~R_{\odot}}$ making the seismic radii slightly overestimated compared to the \textit{Gaia} radii.
They may be slightly more overestimated than this number suggests if the \textit{Gaia} radii are in fact underestimated as discussed above.

In \autoref{fig:rad_ratio} the ratios of seismic to \textit{Gaia} radii are shown as a function of both $T_{\mathrm{eff}}$ and [Fe/H].
The overestimation of the seismic radii based on scaling relations, and the slope in \autoref{fig:rad_comp}, is mainly due to an offset at the highest temperatures in the sample where the mean offset reaches about 7 per cent.
For radii based on individual frequencies, the offset is more or less constant as a function of both $T_{\mathrm{eff}}$ and [Fe/H] with a mean ratio of ${\langle R_{\mathrm{seismic}}/R_{\mathrm{Gaia}}\rangle = 0.988 \pm 0.004}$.
Again, this offset is consistent with a negative bias in the \textit{Gaia} parallaxes of about $30$~$\mu$as, indicating that frequency fitting gives radii accurate to within 1 per cent.
Furthermore, this confirms that the offset of 3 per cent between seismic and \textit{Gaia} parallaxes found previously \citep{2017ApJ...835..173S,2018MNRAS.476.1931S} was mainly caused by a bias in the \textit{Gaia} DR1 parallaxes for these particular stars.

The fact that the offset in the radii from the scaling relations depends on the temperature can be explained by biases in the scaling relations.
Comparisons have been made for stellar models between $\Delta\nu$ from the scaling relation (\autoref{eq:dnu_scal}) and $\Delta\nu$ from theoretical oscillations frequencies of the models \citep[e.g][]{2011ApJ...743..161W}.
They show that the size of the deviation between the two is mainly a function of temperature.
To investigate whether a correction to the scaling relation for $\Delta\nu$ can explain the temperature trend we see, we have derived correction factors for $\Delta\nu$, $f_{\Delta\nu}$.
For each star we take the best-fitting model from the individual frequency fits and calculate the large frequency separation based on the model frequencies following \citet{2011ApJ...743..161W}; this value is referred to as $\Delta\nu_{\mathrm{fit}}$.
Using the mass, radius, and $T_{\mathrm{eff}}$ of the same model, we also calculate $\Delta\nu_{\mathrm{scal}}$ using \autoref{eq:dnu_scal} and define ${f_{\Delta\nu} = \Delta\nu_{\mathrm{fit}} / \Delta\nu_{\mathrm{scal}}}$.
We find that our corrections agree qualitatively with what others have found when comparing $\Delta\nu_{\mathrm{fit}}$ and $\Delta\nu_{\mathrm{scal}}$ for stellar models \citep{2011ApJ...743..161W,2016ApJ...822...15S, 2017ApJS..233...23S}.

Due to the previously mentioned surface effect which causes the high-order model frequencies to be overestimated, $\Delta\nu_{\mathrm{fit}}$ is also overestimated if this effect is not taken into account.
Therefore, we calculate a second set of $\Delta\nu_{\mathrm{fit}}$ and $f_{\Delta\nu}$ after applying the \citet{2014A&A...568A.123B} correction for the surface effect to the model frequencies.

\autoref{fig:rad_ratio_fit}a shows the ratios of seismic to \textit{Gaia} radii with the uncorrected scaling relations (which is the data behind the bins in the upper panel of \autoref{fig:rad_ratio}), and \autoref{fig:rad_ratio_fit}b,c show the same after applying the correction factor by dividing $\Delta\nu$ by $f_{\Delta\nu}$ in \autoref{eq:R_scal}.
After correction of $\Delta\nu$ based on $\Delta\nu_{\mathrm{fit}}$ without correcting for the surface effect, the temperature trend is reduced but only by increasing the radii of stars at intermediate temperatures which leaves a significant mean offset.
In fact, the agreement between the seismic and \textit{Gaia} radii is significantly worse for stars of intermediate temperatures after applying this correction.
For these stars, the good agreement is restored when we take into account the surface effect as seen in \autoref{fig:rad_ratio_fit}c, which also brings the offset down to about 3 per cent at the highest temperatures.

Overall, these results show that theoretically motivated corrections to $\Delta\nu$ improve the accuracy of the seismic radii of dwarfs from scaling relations, but it is important to take the surface effect into account.
After correction including the surface effect, the radii from the scaling relations agree with the \textit{Gaia} radii within 2--3 per cent across the temperature range considered here.
In this work we have calculated a correction for the surface effect for each individual star.
This is possible since we have the observed frequencies for all of them, but this is not generally the case when applying the scaling relations to larger samples of stars.
However, we find that the impact of the surface effect on the correction factors, $f_{\Delta\nu}$, is very close to being a constant offset.
Therefore, it is a good approximation to simply apply the surface correction obtained for the Sun to all dwarfs in the temperature range 5400--6700~K.
This can be done by scaling all values of $f_{\Delta\nu}$ with a common factor such that the value is $1.0$ for a solar model.

Although the temperature trend is reduced by the $\Delta\nu$ correction, it is not completely gone in \autoref{fig:rad_ratio_fit}c.
The remaining temperature trend could be due to inaccuracies in the scaling relation for $\nu_{\mathrm{max}}$ which we have not attempted to include here.

\section{Conclusions}
We have computed seismic radii based on individual frequency modelling and the scaling relations for 93 dwarf stars observed by \textit{Kepler} and included in \textit{Gaia} DR2.
By combining the \textit{Gaia} parallaxes with IRFM angular diameters, we have computed \textit{Gaia} radii for the stars and used them to test the accuracy of the seismic radii.
After removing nine $3\sigma$ outliers, we find only small, although statistically significant, differences between seismic and \textit{Gaia} radii.
The mean radius differences, in the sense seismic minus \textit{Gaia}, are $-0.015~R_{\odot}$ and $0.014~R_{\odot}$ for individual frequencies and scaling relations, respectively.

The offset of the radii based on individual frequencies is equivalent to an offset in parallax of ${\langle\varpi_{\mathrm{Gaia}}-\varpi_{\mathrm{seismic}}\rangle = -35\pm 16~\mathrm{\upmu as}}$.
Since this is consistent with the possible negative bias in the \textit{Gaia} DR2 parallaxes, the bias may be in the \textit{Gaia} radii rather than the seismic ones.
With this in mind, the radii from individual frequencies are likely accurate to a level of 1 per cent or better.

We find that the scaling relations overestimate the stellar radii mainly at high temperatures, reaching values of over 5 per cent larger than the \textit{Gaia} radii.
By introducing theoretically motivated correction factors to $\Delta\nu$ based on individual frequencies of stellar models, we are able to reduce this offset.
However, the corrections actually worsen the accuracy of the scaling relations unless we correct the model frequencies for the surface effect.
After including the surface effect, our corrections bring the seismic radii into average agreement with the \textit{Gaia} radii, and the average offset at the highest temperatures is brought down to 3 per cent.
Taking into account the possible negative bias in the \textit{Gaia} parallaxes, we conclude that the corrected scaling relation for the radius is accurate to within about 2--3 per cent for dwarfs.

\section*{Acknowledgements}
C.L.S. was supported by the project grant `The New Milky Way' from the Knut and Alice Wallenberg foundation.
Funding for the Stellar Astrophysics Centre is provided by The Danish National Research Foundation (Grant agreement No.~DNRF106).
V.S.A. acknowledges support from VILLUM FONDEN (research grant 10118).
This work has made use of data from the European Space Agency (ESA) mission
{\it Gaia} (\url{https://www.cosmos.esa.int/gaia}), processed by the {\it Gaia}
Data Processing and Analysis Consortium (DPAC,
\url{https://www.cosmos.esa.int/web/gaia/dpac/consortium}). Funding for the DPAC
has been provided by national institutions, in particular the institutions
participating in the {\it Gaia} Multilateral Agreement.




\bibliographystyle{mnras}
\bibliography{references}







\bsp	
\label{lastpage}
\end{document}